# Observations of coherence de Broglie waves

Byoung S. Ham

Center for Photon Information Processing, School of Electrical Engineering and Computer Science, Gwangju Institute of Science and Technology

123 Chumdangwagi-ro, Buk-gu, Gwangju 61005, S. Korea

(Submitted on July 10, 2020: bham@gist.ac.kr)

Photonic de Broglie waves (PBWs) via two-mode entangled photon pair interactions on a beam splitter show a pure quantum feature which cannot be obtained by classical means[1-4]. Although PBWs have been intensively studied for quantum metrology[5-13] and quantum sensing[14-25] over the last several decades, their implementation has been limited due to difficulties of high-order NOON state generation[4]. Recently a coherence version of PBWs, the so-called coherence de Broglie waves (CBWs), has been proposed in a pure classical regime of an asymmetrically coupled Mach-Zehnder interferometer (MZI)[26]. Unlike PBWs, the quantumness of CBWs originates from the cascaded quantum superposition of the coupled MZI. Here, the first CBWs observation is presented in a pure classical regime and discussed for its potential applications in coherence quantum metrology to overcome conventional PBWs limited by higher-order entangled photons. To understand the quantum superposition-based nonclassical features in CBWs, various violation tests are also performed, where asymmetrical phase coupling is the key parameter for CBWs.

Quantum metrology[5-13] and quantum sensing[14-25] have been studied over the last several decades to overcome the classical limit in measurement sensitivity and imaging resolution, where quantum measurement error limited by the Heisenberg limit has a square root gain over its classical counterpart[8]. Recently, the advantages of quantum sensing have been demonstrated in various fields for potential applications in geodesy[9,10], lithography[11], imaging[12,13], and magnetometers[14,15]. Although quantum metrology and quantum sensing have great advantages over their classical counterparts, their implementations have been severely limited by the entanglement sources of light[1-4] and matter[21-25]. So far, the largest entangled photon number observed is N=18 for a NOON state generated from an array of quantum dots with a large number of optical devices such as mirrors and beam splitters[4]. Like quantum supremacy in the entangled qubit scale, quantum sensing also has a crossing point with N~100 to achieve practical advantages over its classical counterparts[27]. Such an N for entangled photon numbers may not be possible with current quantum technologies.

Very recently, coherence interpretations have been performed for various quantum features such as anti-correlation on a beam splitter (BS)[28] and unconditional security in key distributions[29], in which quantum superposition plays a major role for the generation of nonclassical phenomena. Quantum entanglement has been conventionally understood as a nonlocal quantum phenomenon that cannot be achieved by classical means[30,31]. According to recent research, however, a quantum feature can also be obtained using coherence optics on a BS if the phase difference between two input fields is set at $\pi/2$[28]. In other words, the nonlclassical features can be a special case of the coherence optics in terms of maximal coherence. Thus, conventional coherence optics can be applied for quantum phenomena. As a result, Mach-Zehnder interferometer (MZI) has been newly interpreted as a quantum device based on maximal coherence in which the $\pi/2$ phase difference induced by the first BS becomes the bedrock of the quantum features such as anticorrelation[28] and nonlocal correlation[32] on the second BS. The propagation directionality in MZI for coherent light may be understood as a macroscopic quantum feature.

The first application of MZI for the nonclassical features is the unconditionally secured classical key distribution (USCKD) in a pure classical regime[29]. For USCKD, a time-reversed unitary transformation is realized in a symmetrically coupled double MZI system. As mentioned above, a quantum feature can be achieved by maximal coherence between two superposed light fields, where the maximal coherence is based on quantum superposition between orthogonal bases[28]. The geometrical structure and mechanism of USCKD is completely different from that of Franson-type experiments[33,34] or energy-time bin-based QKD[35,36]. In a single MZI, there are two orthonormal bases, 0 and $\pi$, as analyzed on a BS[28,29]. In USCKD, a synchronized phase control for the coupled MZIs causes deterministic randomness in the key distribution, where determinacy is used for the key distribution between two remote parties, while randomness is used for unconditional security from an eavesdropper[29]. Thus, the unconditional security of USCKD can be achieved in a purely classical regime and opens the door to macroscopic quantumness.

The second quantum feature based on coherence optics has been presented in the name of coherence de Broglie waves (CBWs) in an asymmetrically coupled double (ACD) MZI system[26]. Unlike entangled photon-based photonic de Broglie waves (PBWs)[1-4], the phase resolution or sensitivity is inversely proportional to N, where N indicates the entangled photon number in a NOON state[1-4]. Thus, the mechanism of CBWs is quite different from PBWs, in which higher-order quantum superposition among coupled MZIs is the physical origin of quantumness. With a chain connection of ACD-MZIs, thus, CBWs can be used for potential applications in coherence quantum technologies of metrology, sensing, geodesy[5,6], imaging[8,9], and inertial navigation[37-39] to overcome the limitations of PBWs[1-4]. Here, we experimentally demonstrate CBWs[26] for the first time and discuss the quantum nature of coherence optics.

### Experimental setup

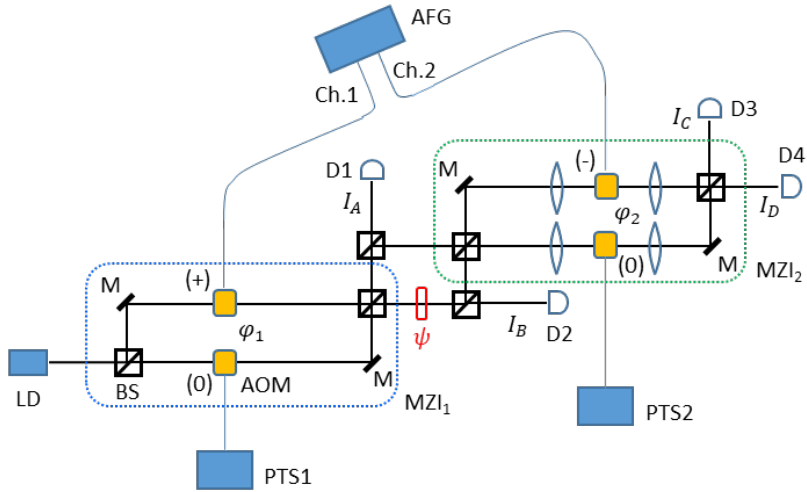

**Fig. 1| An experimental set-up for CBW.** LD: laser diode, BS: unpolarized 50/50 beam splitter, AOM (yellow box): Acousto-optic modulator, PTS: rf generator, M: Mirror, AFG: Arbitrary function generator, D: photo detector. The phase difference of $\varphi_1$ and $\varphi_2$ is due to the frequency difference between two AOMs, where the upper (+)/(-) stands for a positive/negative frequency with respect to (0) at 80 MHz applied to the lower AOMs.

Figure 1 shows a schematic of the ACD-MZI for CBWs, in which two pairs of acousto-optic modulators (AOMs) are synchronously controlled by rf generators, where the AOMs (0) in the lower path are applied by the same frequency as a reference, while the AOMs (+,-) in the upper path are controlled different frequencies to produce $\pm\delta f$ frequency differences. For this, the lower AOMs are applied by 80 MHz rf frequency by both PTS160 and PTS250, while the upper AOMs are applied by 80,000,001 Hz (+) and 79,999,999 Hz (-) to maintain $\delta f = \pm 1\ Hz$, respectively, using a two-channel Tektronix AFG3102. Those four rf fields are synchronized to give the same initial phase $\varphi_0$ to all four AOMs. The physical path length of each MZI is fixed unless specified. Thus, the phase $\pm\delta\varphi$ of each MZI is time-dependent, where $\delta\varphi = \delta f t$. As a result, the light fields, $I_A$ and $I_B$, from the first MZI results in moving fringes whose modulation frequency (period) is 1 Hz (1 s, $\lambda_0$ or $2\pi$), where $\lambda_0$ is the wavelength of the input light $I_0$.

*Theory and numerical analysis*

For theoretical analyses, interference fringes of each output field are calculated. Using MZI matrix representations, the output fields' amplitudes, $E_A$ and $E_B$, from the first MZI in Fig. 1 are as follows:

$$\begin{bmatrix} E_A \\ E_B \end{bmatrix} = [\psi][MZI]_1 \begin{bmatrix} E_0 \\ 0 \end{bmatrix},$$

$$= \frac{1}{2}\begin{bmatrix} 1 & 0 \\ 0 & e^{i\psi} \end{bmatrix}\begin{bmatrix} 1 - e^{i\varphi_1} & i(1 + e^{i\varphi_1}) \\ i(1 + e^{i\varphi_1}) & -(1 - e^{i\varphi_1}) \end{bmatrix}\begin{bmatrix} E_0 \\ 0 \end{bmatrix},$$

$$= \frac{1}{2}\begin{bmatrix} 1 - e^{i\varphi_1} & i(1 + e^{i\varphi_1}) \\ ie^{i\psi}(1 + e^{i\varphi_1}) & -e^{i\psi}(1 - e^{i\varphi_1}) \end{bmatrix}\begin{bmatrix} E_0 \\ 0 \end{bmatrix}.$$

where $[MZI]_1 = [BS][\varphi_1][BS]$, $[\psi] = \begin{bmatrix} 1 & 0 \\ 0 & e^{i\psi} \end{bmatrix}$, $[\varphi_1] = \begin{bmatrix} 1 & 0 \\ 0 & e^{i\varphi_1} \end{bmatrix}$, $[BS] = \frac{1}{\sqrt{2}}\begin{bmatrix} 1 & i \\ i & 1 \end{bmatrix}$, and $E_0$ is the amplitude of the input coherent light from LD (Toptica AT-SHG pro). Thus, the corresponding light intensities $I_A$ and $I_B$ are, respectively:

$$I_A = \frac{I_0}{2}(1 - \cos\delta\varphi), \tag{2}$$

$$I_B = \frac{I_0}{2}(1 + \cos\delta\varphi), \tag{3}$$

where $I_0 = E_0 E_0^*$, $\varphi_1 \equiv \delta\varphi$, and $\delta\varphi$ is the phase difference between two paths of the first MZI. Equations (2) and (3) represent the MZI directionality as a direct result of coherence optics (see the Supplementary Information). Here, the intensity modulation period of $I_A$ and $I_B$ is $2\pi$ or $\lambda_0$, where the phase resolution of $\pi$ (or $\lambda_0/2$) is due to the diffraction limit of classical physics (see the Supplementary Information).

From equations (2) and (3), the final outputs of ACD-MZI in Fig. 1 are calculated as follows (see the Supplementary Information):

$$\begin{bmatrix} E_C \\ E_D \end{bmatrix} = [MZI]_2[\psi][MZI]_1\begin{bmatrix} E_0 \\ 0 \end{bmatrix} =$$

$$\frac{1}{4}\begin{bmatrix} (1 - e^{i\varphi_2})(1 - e^{i\varphi_1}) - e^{i\psi}(1 + e^{i\varphi_1})(1 + e^{i\varphi_1}) & i(1 - e^{i\varphi_2})(1 + e^{i\varphi_1}) - ie^{i\psi}(1 + e^{i\varphi_2})(1 - e^{i\varphi_1}) \\ i(1 + e^{i\varphi_2})(1 - e^{i\varphi_1}) - ie^{i\psi}(1 - e^{i\varphi_2})(1 + e^{i\varphi_1}) & -(1 + e^{i\varphi_2})(1 + e^{i\varphi_1}) + e^{i\psi}(1 - e^{i\varphi_2})(1 - e^{i\varphi_1}) \end{bmatrix}\begin{bmatrix} E_0 \\ 0 \end{bmatrix},$$

where $[MZI]_2 = [BS][\varphi_2][BS]$. For a distinct analysis of Fig. 1, we investigate the output fields for $\pm\delta f$ and $\psi \in \{0, \pi\}$.

(i)      $\varphi_1 = -\varphi_2 = \delta\varphi$ & $\psi = 2n\pi$ (0)

$\begin{bmatrix} E_C \\ E_D \end{bmatrix} = (-1)\begin{bmatrix} cos\delta\varphi & sin\delta\varphi \\ -sin\delta\varphi & cos\delta\varphi \end{bmatrix}\begin{bmatrix} E_0 \\ 0 \end{bmatrix}$ (see the Supplementary Information). For this case, thus, the output intensities are as follows:

$$I_C = \frac{I_0}{2}(1 + cos2\delta\varphi), \tag{4}$$

$$I_D = \frac{I_0}{2}(1 - cos2\delta\varphi), \tag{5}$$

where $\varphi_1$ ($\varphi_2$) represents $+\delta ft$ ($-\delta ft$). Compared with equations (2) and (3), the phase resolution in equations (4) and (5) is doubly enhanced by $\lambda_0/4$, where its effective wavelength is also cut in half at $\lambda_{CBW} = \lambda_0/2$ (see Fig. 2).

Equations (4) and (5) demonstrate the quantum nature of CBW, where $\lambda_{CBW}$ violates the classical physics governed by the diffraction limit of the Rayleigh criterion[26]. Here, the control phase bases of ψ between two MZIs play a critical role in the creation of quantum features based on ordered superposition. As analyzed later, the phase bases of ψ can be replaced by $\pm\delta f$. In other words, the quantum features of CBWs depend on the phase basis choice of ψ and the sign of $\delta f$.

(ii)      $\varphi_1 = -\varphi_2 = \delta\varphi$ & ψ $= (2n-1)\pi\,(\pi)$

$\begin{bmatrix} E_C \\ E_D \end{bmatrix} = \begin{bmatrix} 1 & 0 \\ 0 & -1 \end{bmatrix} \begin{bmatrix} E_0 \\ 0 \end{bmatrix}$. Thus, $I_C = I_0$ and $I_D = 0$ (see Figs. 2c and d). This is the case of USCKD for $\varphi_1 = \varphi_2 = \delta\varphi$ with ψ $= 2n\pi$ [26]. For the opposite cases of (i) and (ii), the sign relation between $\varphi_1$ and $\varphi_2$ is exactly compensated for by the phase basis of ψ in an asymmetrical manner (see the Supplementary Information).

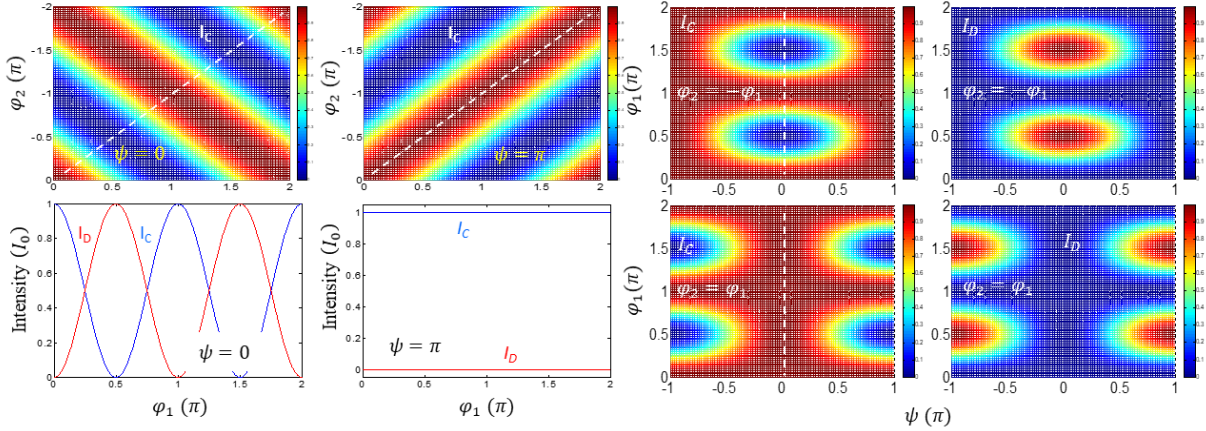

**Fig. 2| Numerical calculations for Fig. 1.** Intensities $I_C$ and $I_D$ as functions of $\varphi_1$, $\varphi_2$, and ψ $= 0$. (first column) CBW; (second column) USCKD; (third column). The dashed line in the upper panels are for lower panels, respectively. (third/fourth column) $\pm\delta f$ −dependent toggle switching.

Figure 2 shows the numerical calculations of CBWs in Fig. 1, where the control of the $\pm\delta f$ (for a fixed ψ) or the coupling phase ψ (for a fixed $\delta f$) acts as a toggle switch between CBW[26] and USCKD[29]. Considering the output intensity from a single MZI, e.g., $I_A$ or $I_B$, whose phase modulation period is $2\pi$ or $\lambda_0$ (see Fig. S2 in the Supplementary Information), Fig. 2 demonstrates the nonclassical features of CBWs at a half-cut modulation period of $\pi$ or $\lambda_0/2$, resulting in a doubled modulation frequency at 2 Hz (see the lower left panel). If the coupling phase is switched from ψ $= 0$ to ψ $= \pi$ for $+\delta f$ to $\varphi_1$ and $-\delta f$ to $\varphi_2$ satisfying ACD-MZI, the moving frame turns out to be frozen, resulting in an identity relation between the input ($I_0$) and output ($I_C$) (see the dashed lines and lower panels). The same effect is also accomplished by controlling $\delta f$ for a fixed ψ (see the third and fourth columns). Thus, the role of $\pm\delta f$ and ψ is same for toggle switching between two eigenvalues (discussed in Discussion).

### Experimental results: CBW

For the experimental demonstrations of CBW, we use a frequency control of $\pm\delta f$ for fixed ψ $= 0$ instead of a path-length control. Thus, the output fringe becomes time  dependent, exhibiting a moving frame: $\delta\varphi = \delta f t$. In each MZI of Fig. 1, the frequency offset $\delta f$ is controlled by the driving frequency for AOM pairs. For $\delta f = +1\,Hz$, thus, the modulation speed of the moving frame in CBW ($I_C$ and $I_D$) is expected to be 2 Hz, as shown in Fig. 2.

Figure 3 shows experimental data of CBWs corresponding to Fig. 2, where the half modulation period of $\pi$ ($\lambda_0/2$) is achieved for the ACD-MZI of Fig. 1. Figure 3a shows CBW with intended violations (see the shaded area) for comparison purposes. With blockage of the ψ −path in Fig. 1 at t~4 s (see the left green arrow in Fig. 3a), the modulation period returns to the normal case of conventional MZI at 1 Hz, resulting in quantum feature demolition. By opening the $\psi$ −path blockage at t~8 s (see the right green arrow), the 2 Hz modulation frequency of CBW is retrieved. The observed CBW in the output intensity $I_C$ has the opposite pattern with respect to $I_D$, as shown in Figs. 3b~d (see the red curves).

In Fig. 3b, another CBW violation test is performed by blocking one of the lower (reference) paths of ACD-MZI, where the shaded area A (B) is with the lower path blockage in the first (second) MZI of Fig. 1. The different intensities in the shaded area are due to unbalanced MZI caused by different efficiencies of AOMs. In short, the doubled phase modulation speed at 2 $Hz$ observed in Figs. 3a and b is a direct proof of the nonclassical features of CBWs overcoming the diffraction limit of the Rayleigh criterion given by equations (2) and (3). This is the novel quantum behavior of ACD-MZI for CBWs. As analyzed in ref. 26, the phase modulation ($n\delta\varphi$) in the output fields increases linearly as the number (n) of ACD-MZI increases. Thus, the ACD-MZI scheme of Fig. 1 becomes the basic building block of the present CBW whose effective wavelength is $\lambda_{CBW} = \lambda_0/2n$. On the contrary, the effective wavelength of PBWs is $\lambda_{PBW} = \lambda_0/2N$. Here, it should be noted that $\lambda_{CBW}$ is for $g^{(1)}$, while $\lambda_{PBW}$ is for $g^{(2)}$.

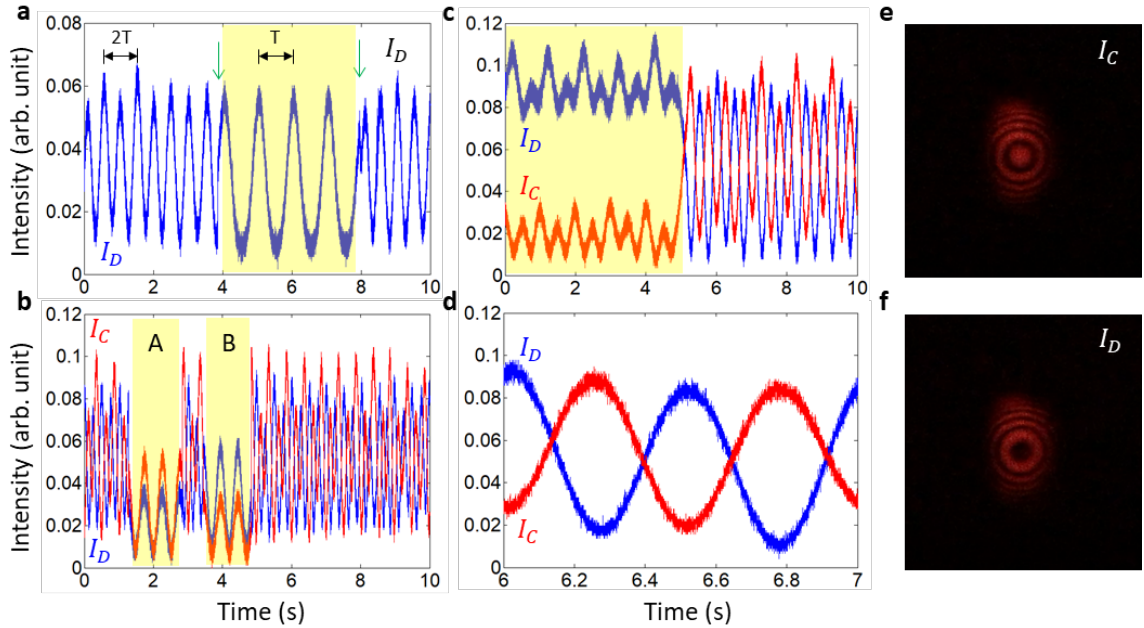

**Fig. 3| Output intensities of Fig. 1 for CBW. a,** Output intensity $I_D$ for CBW. The shaded area is for non-CBW by blocking the $\psi$ −path of the intermediate MZI. The green arrow indicates switching time. T represents a modulation period. **b,** CBW for both $I_C$ and $I_D$. The shaded area is non-CBW by blocking the reference path of the first MZI (A) and second MZI (B). **c,** The $\pm\delta f$ − dependent toggle switching between USCKD and CBW. The shaded area is for USCKD. **d,** Extended data of (**c**) for CBW. **e,f,** Respective fringe patterns of $I_C$ and $I_D$ for CBW at an arbitrary time for maximum and minimum.

Figure 3c shows swapping between CBW[26] and USCKD[29] as shown with the dashed lines in Fig. 2, where the swapping occurs if the sign of $\delta f$ for $\varphi_2$ (AOM, −) is reversed for a fixed $\varphi_1$ (AOM, +), and vice versa (see Figs. S7~S9 in the Supplementary Information). In other words, the AOM driving frequency for Ch.2 in Fig. 1 is switched from 80,000,0001 Hz to 79,999,999 Hz for a fixed AOM frequency of 80,000,001 Hz for Ch.1. The lower AOM frequency is always set at sharp 80 MHz. As calculated in Fig. 2, this identity relation between the input and output is due to a time-reversed unitary transformation (see the shaded area)[29]. Here, the eigenvalues (0, 1) of $I_C$ and $I_D$ can be swapped if $\delta\varphi \to \delta\varphi \pm \pi$ (see Fig. S10 in the Supplementary Information). Because the sign reversal in $\delta f$ affects $\delta\varphi$ in the same way as $\psi$ does, the control phase $\psi$ can replace the $\delta f$ control function. In other words, the same toggle switching between CBWs and USCKD is obtained by an alternative basis choice of $\psi$, either 0 or $\pi$, for a fixed $\pm\delta f$ configuration (see Fig. S11 in the Supplementary Information). Here, the wiggling in the shaded area is due to imperfect MZI caused by misaligned light overlap, phase mismatching caused by air fluctuation, or imperfect $\psi$ basis choice. Figure 3d is a partial extension of Fig. 3c for CBWs, where the output intensities of $I_C$ and $I_D$ are out-of-phase with the modulation period of 0.5 s.

Figures 3e and f show 2D images of the outputs $I_C$ and $I_D$ for CBWs simultaneously taken at the same arbitrary time on both screens, which correspond to Fig. 3d. The Newton's ring-like fringe pattern of $I_C$ and $I_D$ is due to the collimating lens pair across each AOM in the $\varphi_2 -$MZI (see Fig. 1). On the contrary, no lens is used for the $\varphi_1 -$MZI, resulting in a bar fringe (see Fig. S12 in the Supplementary Information).

Figure 4 shows movie files of CBWs recorded for $I_D$ corresponding to Figs. 3a and c. The upper row is captured images of CBW moving frames showing 2 Hz modulation as presented in Figs. 2 and 3. The violation of the CBW in the lower row is viewed by blocking the $\psi -$path, resulting in 1 Hz moving frame speed. Thus, Fig. 4 demonstrates the nonclassical 2D feature of CBWs over the classical limit. In the attached movie file, the first half of the record is for Fig. 3a, and the second half is for Fig. 3c. Due to air fluctuations and the imperfect $\psi -$based phase, there are some phase jitters in the movie frames.

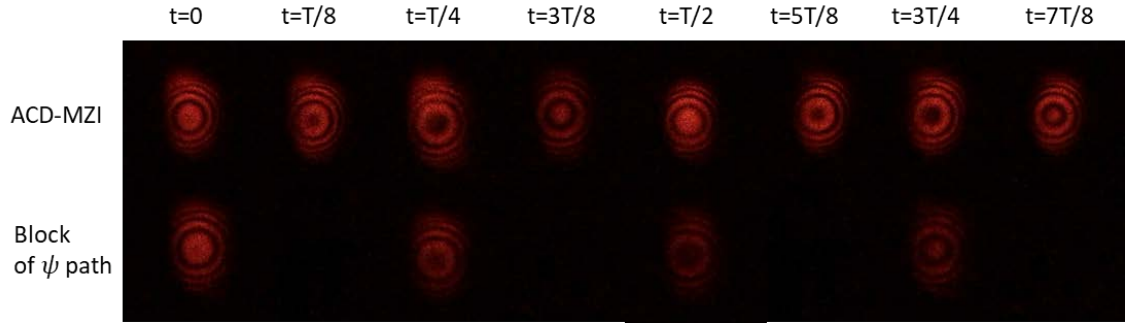

fig\cbw-usckd IMG_0037.MOVD:\논문\2020\논문\cbw 06\fig\cbw-usckd IMG_0038.MOV

**Fig. 4| 2D still images of $I_D$ for CBW and its control (Fig. 3a).** The upper row is for CBW, while the lower row is for the demolition of CBW by blocking the $\psi -$path in Fig. 1. T is the period of diffraction limit. The below movie file is for Figs. 3a and c.

*Discussion*

The origin of the observed nonclassical features of CBW in Figs. 2~4 is due to the double quantum superposition in an ACD-MZI of Fig. 1. To understand the geometry-based quantum phenomenon in ACD-MZI, it may be helpful to review anticorrelation on a BS[28], where the origin of quantumness is in the specific phase relation between two input photons. This $\pm\pi/2$ phase relation between two impinging photons on a BS results in maximum coherence based on two orthogonal bases of sine and cosine functions in a Hilbert space for the BS matrix. In other words, a nonclassical quantum phenomenon can be driven by using normal bases of a classical system. The $\pm\pi/2$ phase requirement on a BS is, however, automatically achieved in the MZI for any light[28]. Thus, MZI functions as a quantum device regardless of input light characteristics. This is the new interpretation of path superposition in the ACD-MZI scheme for CBWs.

The superposed output light from the first MZI in Fig. 1, e.g., $E_A$ at $\delta\varphi = \pi$, is fed into the next MZI via the control phase $\psi$. The (bunched) output field $E_A$ or $E_B$ from the first MZI is split into two superposed fields for the next MZI whose controllability is either $\pm\delta f$ or $\psi \in \{0, \pi\}$. This is an essential step toward the creation of CBWs in the second MZI. As a result, the output light, e.g., $E_D$ experiences double superposition of equation (5). The double superposition is one of two mode eigenstates, where the other is the identity relation for USCKD:

$$(00, c1c2) = (0, c1) \otimes (0, c2), \tag{6}$$

where c1 and c2 corresponds to each eigenstate of each MZI in Fig. 1. The eigenstates of $\pm\delta f$ is represented by $(0, c1)$, while $\psi$ is by $(0, c2)$. Here, $c1$ and $c2$ correspond to nonzero eigenvalues. Thus, equation (6) can be extended to a n-coupled MZI system satisfying the asymmetrical phase relationship of equations (4) and (5). Here, it should be noted that c1 and c2 values are denoted by exponential functions, in which the phase of each eigenstate $|0\rangle$ and $|c1\rangle$ ($|c2\rangle$) is zero and $\varphi_1$ ($\varphi_2$), respectively. Thus, the eigenvalue of $|c1c2 \ldots cn\rangle$ results in the sum phase of $\varphi_1 + \varphi_2 + \cdots + \varphi_n = n\varphi$. This is the principle of the enhanced resolution in CBW as a nonclassical feature of quantumness observed in Fig. 3.

In a serial connection of ACD-MZI, thus, the output field should experience cascaded (ordered) superposition, resulting in the effective wavelength of $\lambda_{CBW} = \lambda_0/2n$, where n is the number of ACD-MZI. For the CBWs, there is no restriction to the input light whether it is a single photon, entangled photon, or a coherence field. Unlike PBWs based on N-entangled photons, CBWs rely on higher-order quantum superposition via the geometric structure of ACD-MZI. As observed in Figs. 3 and 4, the coupled quantum superposition is simply controlled by the binary basis choice of MZI.

For potential applications of CBW, quantum lithography[11] or inertial navigation[37-39] is a good candidate for quantum sensing, where the benefit is in the unlimited order n for $\lambda_{CBW} = \lambda_0/2n$. Regarding quantum lithography, a bulky MZI system (Fig. 1) can also be miniaturized owing to silicon photonics to completely avoid of air fluctuation-caused phase jitters[12,13]. The effective wavelength of CBW results in a shorter wavelength-caused high resolution image. Regarding inertial navigation, a quantum Sagnac interferometer would be a potential candidate for the present CBW[38], where several orders of the value of n for ACD-MZI has a great potential in ultrahigh precision/positioning system for unmanned vehicles, drones, submarines operating without the help of GPS[39], and gravitational wave detections[40,41].

*Conclusion*

Nonclassical features of the coherence de Broglie waves (CBW) were observed in an asymmetrically coupled Mach-Zehnder interferometer (ACD-MZI) for the first time using frequency offset AOMs, resulting in a moving fringe whose phase modulation period cannot be obtained by any classical means. For the control of CBWs, coupled MZIs are manipulated with the frequency offset $\pm\delta f$ or the control phase $\psi$, where the basis choice plays an essential role for determining nonclassicality. As a result, the fringe resolution in the output light was doubled in the ACD-MZI compared with the conventional MZI limited by Rayleigh criterion. Unlike PBWs based on entangled photon pairs, the doubly enhanced resolving power in CBWs was achieved by entirely classical means comparable to N=4 in a PBW NOON state. Instead of the second-order correlation $g^{(2)}$ for PBWs, CBWs are good enough for the first-order correlation, $g^{(1)}$. To prove the coherence (phase)-dependent quantumness in CBWs, each MZI was tested to violate local superposition, resulting in demolition of the nonclassical features of CBWs. Thus, the observed phase (time)-dependent nonclassical feature in the output fringe directly proves the quantum nature of CBWs. In addition to CBWs, the coherence control of ACD-MZI was demonstrated for on-demand mode transfer between CBW and USCKD, where USCKD is for the identity matrix relation between the input and output fields in a classical domain. Thus, higher-order quantum superposition in a cascaded MZI scheme should provide great benefits regarding CBW applications, where a new realm of coherence quantum metrology is opened for quantum sensing, quantum lithography, quantum imaging including LiDAR, quantum inertial navigation, and gravitational wave detections in a purely classical regime with coherent light.

**Methods**

For the ACD-MZI in Fig. 1, acousto-optic modulators (AOMs) are used for frequency-dependent phase control, where inherently given perfect coherence between two paths of MZI is replaced by an AOM-driven frequency offset $\delta f$ controlled by synchronized rf generators, PTS160, PTS250 and Tektronix AFG3102. Thus, the original path-length dependent CBW in ref. [26] is replaced by an rf-driving frequency-controlled AOM-based CBW scheme. The input coherent light $E_0$ is from the Toptica laser (AT-SHG pro) whose wavelength and bandwidth are 605.966 nm and ~300 kHz, respectively. Due to MZI physics, the laser fluctuations in both intensity and phase do not matter on the present experimental results for lower-order coupled systems with n=1, but may affect higher-order CBWs for n $\gg$ 1. To keep a proper diffraction efficiency in AOM in a lengthy ACD-MZI system, a small diameter beam size is maintained by a pair of convex lenses across each AOM for the second MZI (see Fig. 1). Thus, the fringe pattern of the output light for CBWs is due to the combination of linear and circular interferences (see Fig. 4). Unlike the first MZI fringe, thus, the final MZI output shows a Newton's ring-like pattern (see Figs. 3e and f). The input laser power is not sensitive to CBW, but has been kept at about a few mW for its power. For Figs. 2 and 3, Hamamatsu avalanche photodiodes (C12703) are used to record the data using a Tektronix oscilloscope (DPO5204B). For the data in Fig. 3, an iris is added to pass only the zeroth-order fringe and to focus onto the detector (C12703). For the 2D images and movies in Figs. 3e, 3f, and 4, the output lights of $I_C$ and $I_D$ are shined on a paper screen and the images were captured via iPhones. The frequency offset $\pm\delta f$ between two upper paths of ACD-MZI via AOMs is controlled by a two-channel arbitrary function generator (Tektronix AFG3102), whose frequency resolution is 0.001 Hz. All data in Figs. 3 and 4 are raw single-shot recordings without averaging or trimming. The major error source in data is the air fluctuations in each MZI, whose path length is ~60 cm long uncovered. In other words, the experimental setup is rough, coarse, and noisy even without a dark room environment. Thus, the observed CBW exhibits robust to normal conditions and is good for potential applications of coherence quantum metrology and quantum sensing.

**Acknowledgment:** The author acknowledges that the present work was supported by a GRI grant funded by GIST in 2020.